\documentclass{ifacconf}

\usepackage{graphicx}      
\usepackage{natbib}        

\usepackage{amsmath}
\usepackage{amssymb}
\usepackage{bm}
\usepackage{ifthen}
\usepackage{algorithm}
\usepackage{algpseudocode}
\begin{document}

\newcommand{\est}[1]{\ensuremath{\hat{#1}}}		
\newcommand{\given}{\ensuremath{\!\mid\!}}								
\newcommand{\func}[2]{\ensuremath{#1 \!\left( #2\right)}}				
\newcommand{\dif}[1][d]{\ensuremath{\, \nit{#1}}}						
\newcommand{\lab}[1]{\ensuremath{\tilde{#1}}}							
\newcommand{\sip}[2]{\ensuremath{\left\langle #1, #2 \right\rangle}}	

\newcommand{\nit}[1]{\mathrm{#1}}
\newcommand{\kk}[1][0]{\ifthenelse{\equal{#1}{0}}{\ensuremath{k \mid k}}{\ensuremath{k \mid k-#1}}}

\begin{frontmatter}

\title{A Track-Before-Detect Approach to Multi-Target Tracking on Automotive Radar Sensor Data} 

\author[First]{David Meister} 
\author[Second]{Martin F. Holder} 
\author[Third]{Hermann Winner}

\address[First]{University of Stuttgart, Germany, (e-mail: david.meister@ist.uni-stuttgart.de)}
\address[Second]{TU Darmstadt, Germany, (e-mail: holder@fzd.tu-darmstadt.de)}
\address[Third]{TU Darmstadt, Germany, (e-mail: winner@fzd.tu-darmstadt.de)}

\begin{abstract}                
In recent years, Bayes filter methods in the labeled random finite set formulation have become increasingly powerful in the multi-target tracking domain. 
One of the latest outcomes is the Generalized Labeled Multi-Bernoulli (GLMB) filter which allows for stable cardinality and target state estimation as well as target identification in a unified framework. 
In contrast to the initial context of the GLMB filter, this paper makes use of it in the Track-Before-Detect (TBD) scheme and thus, avoids information loss due to thresholding and other data preprocessing steps. 
This paper provides a TBD GLMB filter design under the separable likelihood assumption that can be applied to real world scenarios and data in the automotive radar context. 
Its applicability to real sensor data is demonstrated in an exemplary scenario.
To the best of the authors' knowledge, the GLMB filter is applied to real radar data in a TBD framework for the first time.
\end{abstract}

\begin{keyword}
Bayesian methods, Particle filtering, Estimation and filtering, Sensor integration and perception, Random finite sets, Multi-object tracking, GLMB, Track-before-detect
\end{keyword}

\end{frontmatter}

\section{Introduction}

Radar sensors are widely used for object detection in the automotive domain due to their ability to simultaneously measure range and relative velocity as well as their robustness against adverse weather conditions.
The measurements of a radar sensor are initially available as a multidimensional spectrum in the dimensions distance, relative radial velocity and azimuth angle.
In a typical radar signal processing chain, an (adaptive) threshold is applied to detect individual targets.
Subsequently, a tracking process relates targets at different time steps.
On the contrary, in the so-called Track-Before-Detect (TBD) framework, objects are estimated directly from the spectral data without any thresholding.
This eliminates potential information loss and has been widely discussed in the literature, e.g.\ by \cite{Boers.2003}, \cite{Vo.2010}, or \cite{Papi.2013}.

In general, multi-target tracking is an estimation problem where targets and their states are to be identified individually.
Multi-target Bayes estimators in the Random Finite Set (RFS) formulation, introduced by \cite{Mahler.2003}, \cite{Vo.2010}, \cite{Vo.2013b}, and others, allow for the simultaneous estimation of the number of objects and their states.
The GLMB filter developed by \cite{Vo.2013b} is one of the latest RFS multi-target tracking approaches and according to \cite{Vo.2015}, the first exact closed form solution to the multi-target Bayes recursion that allows for an unbiased estimation of target number and states as well as the unique identification of each target.
\cite{Papi.2013} apply the TBD idea to the GLMB filter framework from a theoretical perspective.
This paper extends their work to make it applicable to real radar sensor data.

\section{Multi-Target Tracking with the GLMB filter}

This section introduces the GLMB filter recursion equations and the necessary labeled RFS theory.
Let
\begin{equation*}
\func{1_S}{X} = 
\begin{cases}
	1, & X \subseteq S \\
	0, & \text{otherwise}
\end{cases}, \quad
\func{\delta_S}{X} = 
\begin{cases}
	1, & X = S \\
	0, & \text{otherwise}
\end{cases}
\end{equation*}
be the inclusion function and the generalization of the Dirac delta function, respectively.

\subsection{GLMB Random Finite Set}

We introduce the $\delta$-GLMB RFS distribution as the multi-object analogue of a probability density function (PDF) for the set of labeled target state vectors $\lab{X}$:
\begin{equation}
    \lab{\pi}(\lab{X}) = \Delta(\lab{X}) 
    \sum_{\left( I, \xi \right) \in \func{\mathcal{F}}{\mathbb{L}}\times\Xi}
    \omega^{(I, \xi)}
    \delta_{I}(\mathcal{L}(\lab{X}))
    \left[ p^{(\xi)} \right]^{\lab{X}},	\label{eq:d_glmb_pdf_def}
\end{equation}
where $\func{\mathcal{F}}{\mathbb{L}}$ and $\Xi$ denote the space of all finite subsets of the label space $\mathbb{L}$ and a discrete index set, $p^{(\xi)}(\lab{\bm{x}})$ refers to the distribution of a state vector $\lab{\bm{x}}$ in $\lab{X}$, $\omega^{(I, \xi)}$ represents a non-negative weight and a multi-target exponential is defined as $h^{\lab{X}} = \prod_{\lab{\bm{x}} \in \lab{X}} \func{h}{\lab{\bm{x}}}$ with $h^{\emptyset} = 1$.
Furthermore, the distinct label indicator $\Delta(\lab{X}) = \delta_{\lvert \lab{X} \rvert}(\lvert \mathcal{L}(\lab{X}) \rvert)$ compares the cardinality of the RFS $\lab{X}$ with the one of its labels $\mathcal{L}(\lab{X}) = \lbrace \func{\mathcal{L}}{\lab{\bm{x}}} : \lab{\bm{x}} \in \lab{X} \rbrace$ and therefore, indicates if $\lab{X}$ has distinct labels.
Let $\func{\mathcal{L}}{\lab{\bm{x}}} = \func{\mathcal{L}}{\left( \bm{x}, l \right)} = l$ return the label $l$ of $\lab{\bm{x}}$.
Essentially, the labeled multi-object PDF is described by a weighted mixture of multi-target exponentials $[ p^{(\xi)} ]^{\lab{X}}$.
For more details, see \cite{Vo.2013b}.

\subsection{GLMB Filter Framework}  \label{sec:GLMB_filter}

This subsection presents the GLMB filter recursion in the joint prediction and update version according to \cite{Vo.2017}, slightly modified in favor of the TBD use case according to \cite{Papi.2015} as well as to accommodate the birth model presented in Section~\ref{sec:abm}.
Let $\sip{f}{g} = \int \func{f}{x} \func{g}{x} \dif x$ abbreviate the standard inner product.

The joint prediction and update directly leading from the previous GLMB posterior density
\begin{multline*}
\lab{\pi}_{k-1}(\lab{X}_{k-1} \given \bm{z}_{1:k-1})
\propto
\Delta(\lab{X}_{k-1}) \\
\cdot \sum_{I_{k-1} \in \func{\mathcal{F}}{\mathbb{L}_{0:k-1}}}
\omega_{k-1}^{(I_{k-1})}
\delta_{I_{k-1}}(\mathcal{L}(\lab{X}_{k-1}))
\left[ p_{k-1}^{(I_{k-1})} \right]^{\lab{X}_{k-1}},
\end{multline*}
to the one at the current time step can be expressed as
\begin{multline*}
\lab{\pi}_{k}(\lab{X}_{k} \given \bm{z}_{1:k}) \propto
\Delta(\lab{X}_{k}) \\
\cdot \sum_{\substack{I_{k-1} \in \func{\mathcal{F}}{\mathbb{L}_{0:k-1}},\\
	  I_{k} \in \func{\mathcal{F}}{\mathbb{L}_{0:k}}}}
\omega_{k-1}^{(I_{k-1})}
\omega_{\bm{z}_{k}}^{(I_{k-1}, I_{k})}
\delta_{I_{k}}(\mathcal{L}(\lab{X}_{k}))
\left[ p_{k}^{(I_{k})} \right]^{\lab{X}_{k}},
\label{eq:glmb_jup_post}
\end{multline*}
where the $\delta$-GLMB form introduced in \eqref{eq:d_glmb_pdf_def} has been used and $\bm{z}_{1:k}$ abbreviates the set of measurements received from the first to the $k$-th time step.
Likewise, $\mathbb{L}_{0:k}$ refers to the label space built from the start of the recursion to the $k$-th time step.
Moreover, the following definitions apply:
\begin{align*}
\omega_{\bm{z}_{k}}^{(I_{k-1}, I_{k})} =&
	\left[ 1-\bar{p}_{\nit{S}}^{(I_{k-1})}
		\right]^{I_{k-1}-I_{k}}
	\left[ \bar{p}_{\nit{S}}^{(I_{k-1})} 
		\right]^{I_{k-1} \cap I_{k}} \nonumber\\ \cdot~&
	\left[ 1-\mathfrak{r}_{\nit{B},k}
		\right]^{\func{I_{\nit{B}}}{I_{k-1}} - I_{k}}
	\mathfrak{r}_{\nit{B},k}^{
		\func{I_{\nit{B}}}{I_{k-1}} \cap I_{k}}
	\left[ \bar{\psi}_{\bm{z}_{k}}^{(I_{k})} 
		\right]^{I_{k}}, \\
\func{\bar{p}_{\nit{S}}^{(I_{k-1})}}{l} =& 
	\sip{\func{p_{k-1}^{(I_{k-1})}}{\cdot, l}}{
		\func{p_{\nit{S}}}{\cdot, l}}, \\
\func{\bar{\psi}_{\bm{z}_{k}}^{(I_{k})}}{l} =&
	\sip{\func{p_{\kk[1]}^{(I_{k})}}{\cdot, l}}{
		\func{\psi_{\bm{z}_{k}}}{\cdot, l}},\\
\func{p_{\kk[1]}^{(I_{k})}}{\bm{x}_{k}, l} =&
	\func{1_{\mathbb{L}_{0:k-1}}}{l}
	\func{p_{\nit{S}}^{(I_{k})}}{\bm{x}_k, l} + 
	\func{1_{\mathbb{L}_{k}}}{l}
	\func{p_{\nit{B}, k}}{\bm{x}_{k}, l}, \\
\func{p_{\nit{S}}^{(I_{k})}}{\bm{x}_k, l} =&
\frac{\sip{\func{p_{\nit{S}}}{\cdot, l} 
			\func{f_{\kk[1]}}{\lab{\bm{x}}_{k} \given \cdot, l}}{
			\func{p_{k-1}^{(I_{k-1})}}{\cdot, l}}}{
		\func{\bar{p}_{\nit{S}}^{(I_{k-1})}}{l}},\\
\func{p_{k}^{(I_{k})}}{\bm{x}_{k}, l} =&
	\frac{\func{p_{\kk[1]}^{(I_{k})}}{\bm{x}_{k}, l}
		\func{\psi_{\bm{z}_{k}}}{\bm{x}_{k}, l}}{
		\func{\bar{\psi}_{\bm{z}_{k}}^{(I_{k})}}{l}},
\end{align*}
with $I_i - I_j = \left(I_i \cup I_j\right) \backslash I_j$ and $\func{f_{\kk[1]}}{\lab{\bm{x}}_{k} \given \lab{\bm{x}}_{k-1}}$ referring to the Markov transition density imposed by the transition model presented in Section~\ref{sec:tm}.
The set $I_k$ contains all labels of a potential target constellation at time step $k$, also referred to as hypothesis, while $\func{I_{\nit{B}}}{I_{k-1}}$ denotes the label set of potential new born or appearing targets if $I_{k-1}$ is assumed as the previous target constellation.
The motivation for changing the formulation by \cite{Vo.2017} is to allow for target birth taking into account the non-overlapping illumination region assumption leading to separable likelihoods required to derive the measurement model in Section~\ref{sec:mm}.
Moreover, $p_{\nit{S}}(\lab{\bm{x}}_k)$ quantifies the survival probability of a target, i.e.\ the prior probability of continued existence of a target after one time step.
Assuming a Labelled Multi-Bernoulli RFS for describing the birth or target appearing process, $\mathfrak{r}_{\nit{B}}$ and $p_{\nit{B}}(\lab{\bm{x}})$ can be interpreted as birth probability and state distribution of a new born target.
For the TBD measurement model presented in Section~\ref{sec:mm}, the measurement likelihood contribution of a present target can be written as
\begin{equation}
    \func{\psi_{\bm{z}_{k}}}{\lab{\bm{x}}} = 
    \prod_{\iota \in \func{C}{\lab{\bm{x}}}} 
    \ell^{(\iota)}(z^{(\iota)} \given \lab{\bm{x}})
    \label{eq:meas_lik_tgt_contr}
\end{equation}
where $\ell^{(\iota)}(z^{(\iota)} | \lab{\bm{x}})$ is called likelihood ratio.
What is more, $C(\lab{\bm{x}})$ refers to the illumination region of a specific target representing a set of measurement vector elements $\{z^{(\iota)}\}_{\iota \in \func{C}{\lab{\bm{x}}}}$ influenced by the target's presence.

Gibbs sampling is utilized as the truncation method as suggested by \cite{Vo.2017}.

\section{Radar Sensor}

The Fast Chirp Modulation radar sensor used in the experiments is widely employed in the automotive field.
It combines frequency and pulse modulation by transmitting a sequence of linearly increasing frequency ramps (chirps).
By analyzing phase and frequency shifts as well as time delays of incoming waves reflected by surfaces in the field of view, range, radial relative speed and azimuth angle of targets can be computed.
The intensity of the received wave depends on the absorption and reflection properties of the target surface, the distance to the target and multi-path propagation mitigation effects.
By performing a three-dimensional Fast Fourier Transform, the individual wave properties of multiple reflections can be isolated.
\cite{Wintermantel.2009} describes the aforementioned process yielding the data structure referred to as radar cube in detail.
The measurement space is discretized along the dimensions range, radial relative speed and azimuth angle leading to multiple measurement cells.
They are assigned an intensity proportional to the power of the wave potentially received from a target at the respective location in the measurement space.
By stacking these intensity values in a fixed order, the measurement vector $\bm{z}_k$ is obtained.

\section{Experimental Setup and Models}

This section presents the transition and measurement model as well as the employed adaptive birth model and track merging strategy.
A more detailed discussion on all subsections is provided by \cite{Meister.2019}.

\subsection{Transition Model}   \label{sec:tm}

In multi-target tracking literature, it is common to assume a constant velocity transition model as presented by \cite{Vo.2014} or \cite{Papi.2015}.
With the state vector $\bm{x}_{k} = [x_{k}, \dot{x}_{k}, y_{k}, \dot{y}_{k}, \vartheta_{k}]^T$ consisting in planar position and velocity as well as a constant fifth state to be defined later on, the target transition equation can be expressed as
\begin{equation*}
\bm{x}_k = A \bm{x}_{k-1} + \bm{v}_{k-1}, \quad \bm{v}_{k-1} \sim \mathcal{N}(\bm{0},Q)
\end{equation*}
where
\begin{align*}
    A &= \mathrm{diag}(\bar{A}, \bar{A}, 1), & Q &= \mathrm{diag}(\sigma_{\ddot{x}}^2 \bar{Q}, \sigma_{\ddot{y}}^2 \bar{Q}, \sigma_{\dot{\vartheta}}^2 \Delta t^2) \\
    \bar{A} &= \begin{bmatrix} 1 & \Delta t \\ 0 & 1 \end{bmatrix}, & \bar{Q} &= \begin{bmatrix} \frac{\Delta t^4}{4}  & \frac{\Delta t^3}{2}  \\ \frac{\Delta t^3}{2}  & \Delta t^2  \end{bmatrix}
\end{align*}
and $\Delta t$ refers to the time step size.
The process noise $\bm{v}_{k-1}$ can therefore be interpreted as being caused by a random acceleration between two time steps.
Please note that the model can only give reasonable predictions if the rotational speed of the ego vehicle is negligible. 
Otherwise, the latter would need to be estimated, e.g.\ by an ego motion estimator, to compensate for the relative target state transition due to the sensor rotation.
Thus, ego vehicle steering is avoided in the scenario in Section~\ref{sec:results}.

\subsection{TBD Measurement Model}  \label{sec:mm}

An overview on multi-target likelihoods in the TBD framework is provided by \cite{Lepoutre.2016}.
In this paper, we restrict the implementation to the Swerling 1 case with squared modulus measurements provided by the radar sensor.
When assuming separable likelihoods due to non-overlapping illumination regions and neglecting the spatial coherence of the measurements, the likelihood ratio in \eqref{eq:meas_lik_tgt_contr} for the Swerling 1 case and target $i$ can be derived to be 
\begin{equation*}
    \ell(z_{k}^{(\iota)} \given \lab{\bm{x}}_{k,i}) =
    \frac{1}{1 + \func{\varsigma^{(\iota)}}{\lab{\bm{x}}_{k,i}}} 
    \exp \left( \frac{z_{k}^{(\iota)}
    	\func{\varsigma^{(\iota)}}{\lab{\bm{x}}_{k,i}}}{
    	\func{\mu_{z}^{(\iota)}}{\lab{\bm{x}}_{k,i}}} \right)
    \label{eq:sw1_lik_ratio}
\end{equation*}
where 
$
    \func{\varsigma^{(\iota)}}{\lab{\bm{x}}_{k,i}} = 
    \frac{\sigma_{\rho_i}^2}{\sigma_{w}^2}
    \left\lvert \func{h^{(\iota)}}{\lab{\bm{x}}_{k,i}} \right\rvert^2
$
and 
$
    \func{\mu_{z}^{(\iota)}}{\lab{\bm{x}}_{k,i}} = 
    2\sigma_{w}^2 + 2\sigma_{\rho_i}^2 
    \left\lvert \func{h^{(\iota)}}{\lab{\bm{x}}_{k,i}} \right\rvert^2.
    \label{eq:sw1_exp_meas}
$
The function $\func{h^{(\iota)}}{\lab{\bm{x}}_{k,i}}$ returns the value of a point-spread function in measurement cell $\iota$ and $\sigma_{w}^2$ quantifies the measurement noise variance.

In contrast to the suggestion by \cite{Lepoutre.2016}, the following correspondence is established:
\begin{equation*}
    \vartheta_{k,i} = \sigma_{\rho_i}^2 \cdot
    \frac{\func{G}{\lab{\bm{x}}_{k,i}}^2}{\func{r}{\lab{\bm{x}}_{k,i}}^4},
    \label{eq:sw1_fifth_state}
\end{equation*}
where corrections according to the radar range equation (see \cite{Richards.2014}) by the antenna gain $\func{G}{\lab{\bm{x}}_{k,i}}$ and the range $\func{r}{\lab{\bm{x}}_{k,i}}$ are incorporated.

\subsection{Adaptive Birth Model}   \label{sec:abm}

Since the non-overlapping target illumination region assumption is crucial to the performance of the
filter, a birth model is designed that incorporates present targets and their illumination regions in
the birth process.
The idea behind the birth model utilized in this paper is to initialize targets in regions with significant measurements while ensuring that new born targets do not overlap with existing ones in their illumination regions.
The latter are assumed to be of equal ellipsoidal size in the measurement space to achieve simple equations to check for overlapping illumination regions.
For details, see \cite{Meister.2019}.

\subsection{Track Merging Strategy}

The GLMB filter under the non-overlapping illumination region assumption requires a track merging strategy. 
\cite{Vo.2010} and \cite{Mahler.2014} provide helpful insights into adequate methods for other RFS approaches.
For the reason of simplicity, an either-or-relationship is established between overlapping targets in the Gibbs sampling process here:
only one of the targets can survive the update step.
This is achieved by modifying the Gibbs cost matrix rows of overlapping targets adequately as soon as one of them is sampled to exist, i.e.\ the other targets' death probabilities are artificially set to 1.
Further research could establish more sophisticated merging strategies for the TBD GLMB filter including a combination of elimination regions and track merging as suggested by \cite{Suzuki.2018}.

\section{Experimental Results}  \label{sec:results}

A Sequential Monte Carlo approximation according to \cite{Vo.2014} was utilized to obtain the following results.

\subsection{Scenario Description and Filter Tuning}

\begin{figure}[htb]
\begin{center}
\includegraphics[width=8.7cm]{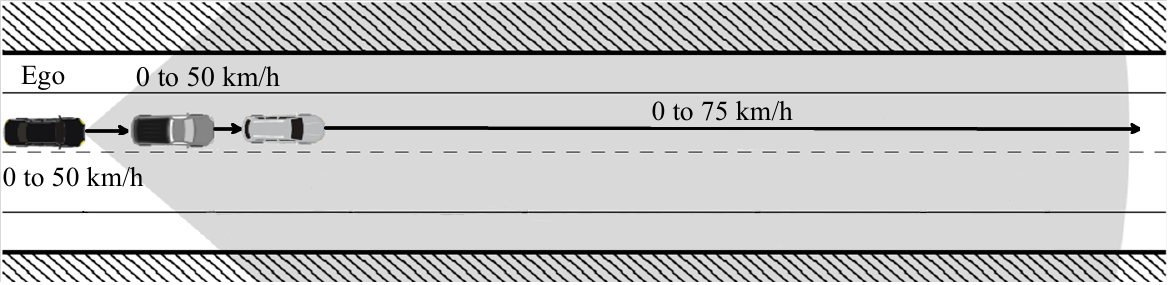}    
\caption{Scenario illustration} 
\label{fig:dyn_scheme}
\end{center}
\end{figure}

The examined scenario incorporates the ego vehicle with the radar sensor mounted at the front and two passenger cars as targets moving in line on a straight road.
The vehicles accelerate to the speeds shown in Fig.~\ref{fig:dyn_scheme}.
The light grey area represents the sensor field of view.
This scenario allows to analyze the filter behavior in a dynamic environment including target occlusion effects while maintaining the transition model assumption of no ego vehicle steering.

The survival probability $p_{\nit{S}}$ is set to $0.99$.
The algorithm considers a maximum of 200 hypotheses and utilizes 15000 particles to represent a target state distribution.
The initial birth existence probability $\mathfrak{r}_{\nit{B}}$ is kept at 0.3 and the illumination region ellipsoid radii are set to the doubled sensor resolution cell size. 
Birth targets are initialized uniformly around the centroid of a measurement cell with a significant intensity above $z_{th} = 10^{-5}$.
The measurement noise $\sigma_{w}^2$ is quantified as $2 \cdot 10^{-6}$.
A new measurement is approximately received every 70~ms.
The process noise variances are $\sigma_{\ddot{x}}^2 = \sigma_{\ddot{y}}^2 = \left(\frac{5}{3}\right)^2~\nit{m}^2\nit{s}^{-4}$ and $\sigma_{\dot{\vartheta}}^2 = 10^{-3}$.

\subsection{Filter Behavior Analysis}

\begin{figure}[ht]
\begin{center}
\includegraphics[width=9cm]{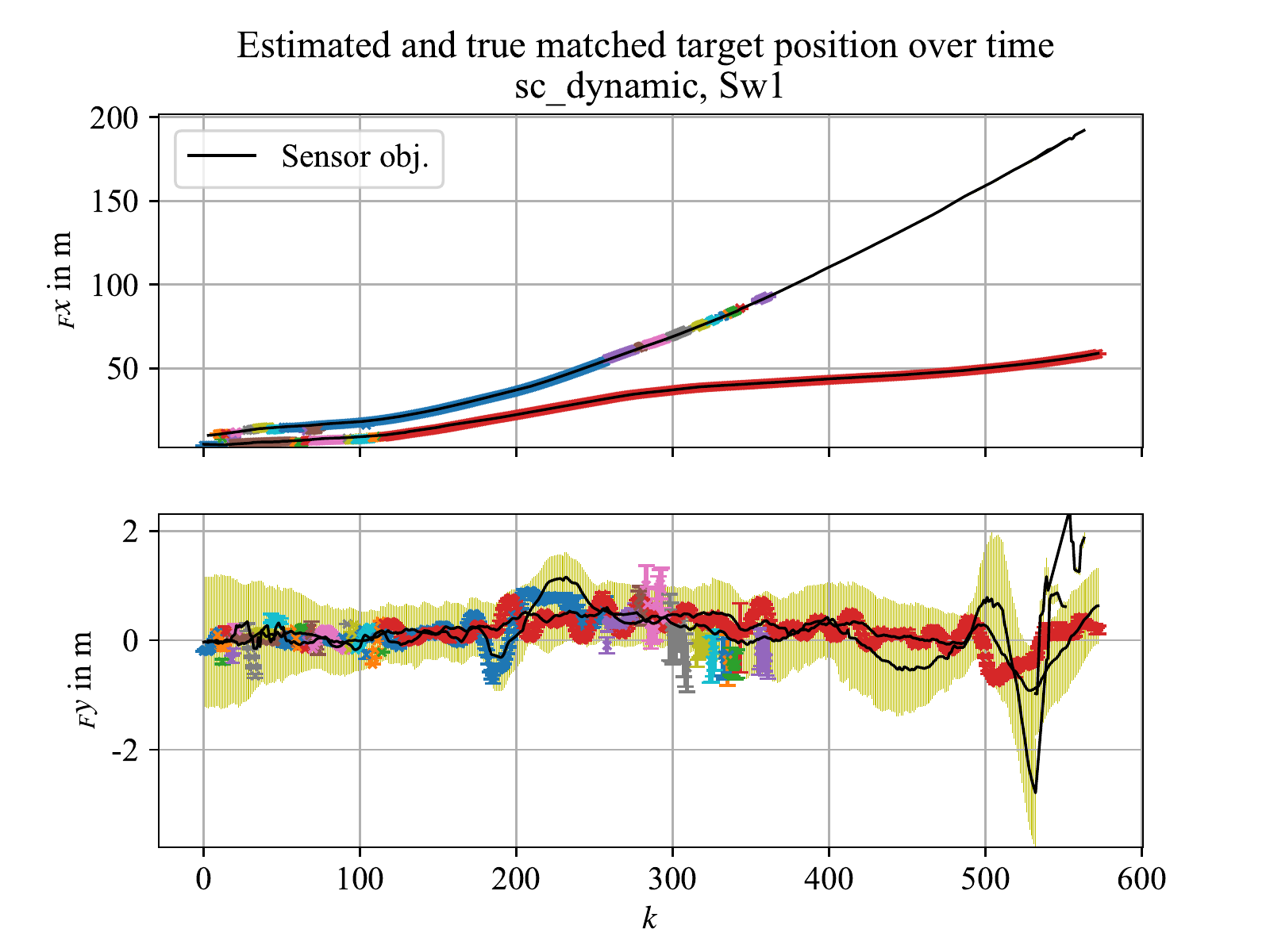}    
\caption{Estimated target positions $x$ and $y$ over time $k$} 
\label{fig:dyn_pos_matched}
\end{center}
\end{figure}

\begin{figure}[ht]
\begin{center}
\includegraphics[width=9cm]{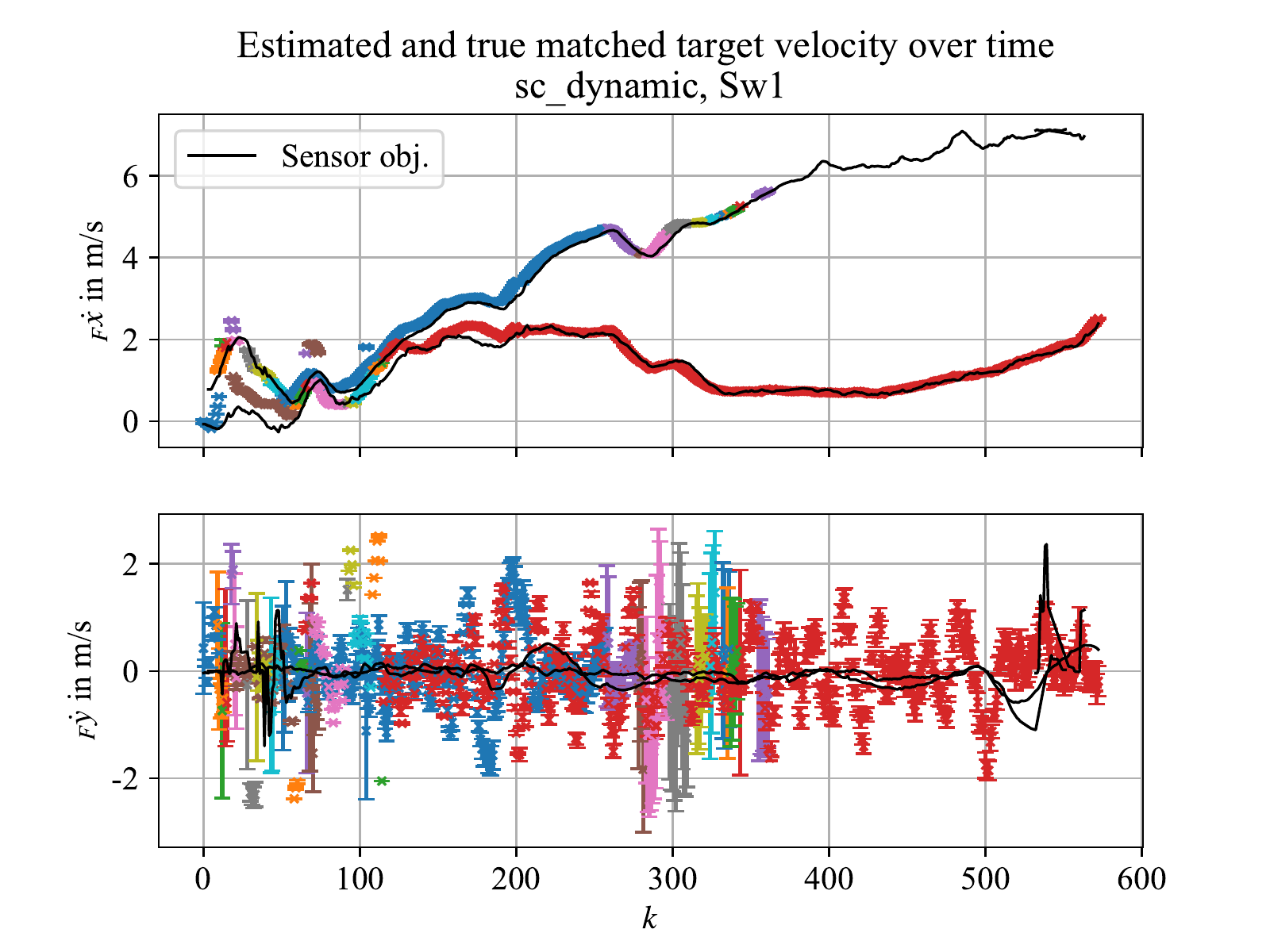}    
\caption{Estimated target velocities $\dot{x}$ and $\dot{y}$ over time $k$} 
\label{fig:dyn_vel_matched}
\end{center}
\end{figure}

The estimated target positions and velocities over time are shown in Fig.~\ref{fig:dyn_pos_matched} and Fig.~\ref{fig:dyn_vel_matched}.
They are depicted along with the estimates of the sensor internal estimation algorithm considered as a ground truth reference.
The coloring of the estimates refers to the unique target identities.
Target estimates at the borders of the field of view have been filtered out due to the fact that they originate from a data ambiguity imposed by the sensor design that was not incorporated in the implemented measurement model.

It can be observed that the implemented TBD GLMB filter estimates the target positions and velocities reasonably:
The lateral distance $y$ to the ego vehicle center stays close to zero at all times.
Small deviations from zero in the lateral direction can be explained by the physical size of the target indicated in yellow.
It is worth highlighting that the filter tracks the centers of reflection and not the target centers.
Moreover, the longitudinal distance $x$ follows the sensor internal estimates shown as black lines very accurately.
Due to multi-path propagation, the sensor can also detect the occluded distant vehicle.
Nonetheless, the filter looses track of the distant vehicle approximately from time step 370 onward.
This is mainly caused by signal mitigation effects for an occluded distant vehicle.
Further tuning of the assumed noise floor might allow for better tracking results like for the sensor internal estimator.

Similar observations can be made for the velocity estimates depicted in Fig.~\ref{fig:dyn_vel_matched}.
The longitudinal velocity estimates follow the radar sensor estimates consistently.
The higher volatility in the targets' lateral speed is induced by the missing measurement information regarding the velocity in that direction.
Moreover, sudden shifts in the reflection center location, potentially induced by slight target orientation variations, result in significant velocity changes.

Lastly, targets are labelled quite persistently which indicates that the TBD GLMB filter is capable of keeping track of target identities in an adequate fashion.

\section{Conclusion}

This paper provides a proof of concept for the implemented TBD GLMB filter with an automotive radar sensor.
A real data experiment has been performed that yields desirable tracking performance. 
A more detailed analysis, also considering more scenarios and elaborating on the made assumptions in depth, is provided by \cite{Meister.2019}.
Further research could focus on more sophisticated track merging strategies or the assessment of different measurement models.
What is more, including weakly reflecting targets in the experiments is expected to allow for a closer evaluation of the magnitude of the mentioned information loss.
Designing a real-time capable TBD GLMB filter algorithm remains an open challenge as well.

\bibliography{references}             
\end{document}